\documentclass[slac_one]{revtex4}
\usepackage{graphicx}
\usepackage{fancyhdr}
\begin{document}

\title{Level-3 Calorimetric Resolution available for the Level-1 and Level-2 CDF Triggers.} 

%

\author{A.Canepa.}
\affiliation{University of Pennsylvania, USA}
\author{M.Casarsa, T.Liu.}
\affiliation{FNAL, Batavia, USA}
\author{G.Cortiana}
\affiliation{Istituto Nazionale di Fisica Nucleare Padova, Italy}

\author{G.Flanagan.}
\affiliation{Purdue University, USA}

\author{H.Frisch, D.Krop, C.Pilcher, V.Rusu.}
\affiliation{University of Chicago, USA}

\author{V.Cavaliere, V.Greco.}
\affiliation{Istituto Nazionale di Fisica Nucleare Siena, Italy}

\author{P.Giannetti, M.Piendibene, L.Sartori.}
\affiliation{Istituto Nazionale di Fisica Nucleare Pisa, Italy}

\author{M.Vidal.}
\affiliation{CIEMAT, Madrid}

\begin{abstract}
As the Tevatron luminosity increases sophisticated selections are required to be efficient in selecting rare events among a very huge background. To cope with this problem, CDF has pushed the offline calorimeter algorithm reconstruction resolution up to Level 2 and, when possible, even up to Level 1, increasing efficiency and, at the same time, keeping under control the rates. 

The CDF Run II Level 2 calorimeter trigger is implemented in hardware and is based on a simple algorithm that was used in Run I. 
This system has worked well for Run II at low luminosity. As the Tevatron instantaneous luminosity increases, the limitation due to this simple algorithm starts to become clear: some of the most important jet and MET (Missing ET) related triggers have large growth terms in cross section at higher luminosity. 
In this paper, we present an upgrade of the Level 2 Calorimeter system which makes the calorimeter trigger tower information available directly to a CPU allowing more sophisticated algorithms to be implemented in software. Both Level 2 jets and MET can be made nearly equivalent to offline quality, thus significantly improving the performance and flexibility of the jet and MET related triggers. 
However in order to fully take advantage of the new L2 triggering capabilities having at Level 1 the same L2 MET resolution is necessary. 
The new Level-1 MET resolution is calculated by dedicated hardware.  
This paper describes the design, the hardware and software implementation and the performance of the upgraded calorimeter trigger system both at Level 2 and Level 1.  
\end{abstract}

\maketitle

\thispagestyle{fancy}

\section{OVERVIEW OF  THE CDF  CALORIMETER TRIGGER} 
The CDF trigger~\cite{TDR} for Run II is a three level system: each stage rejects a sufficient fraction of the events to allow processing at the next 
stage with acceptable dead time. The goal of the calorimeter trigger (both at Level 1 and Level 2) is to trigger on electrons, photons, jets, total transverse energy (SumET) as well as missing transverse energy (MET). In the following we use a coordinate system defined by the polar angle $\theta$ , measured from the proton direction, 
the azimuthal angle $\phi$,  measured from the Tevatron plane. The pseudo-rapidity is defined as $\eta = − \ln(\tan(\theta /2))$.
For CDF Run II, all calorimeter tower energy information, including both electromagnetic (EM) energy and hadronic (HAD) energy, is digitized every $132$ ns 
and the physical towers are summed into trigger towers, weighted by $\sin(\theta)$  to yield transverse energy. A trigger tower covers 15 degree in azimuth $\phi$
and approximately 0.2 in pseudo-rapidity $\eta$. This results in a representation of the entire detector as a $24 \times 24$ map in the $\eta - \phi$ plane. 
The trigger tower energy information is then sent to both L1 and L2 calorimeter trigger systems with 10-bit energy resolution, using a least significant 
count of 125 MeV and resulting in a full scale of 128 GeV.  The Level 1 calorimeter (L1CAL) subsystem only uses 8 of the 10 available bits for each trigger tower, 
with the two least significant bits dropped, giving a least count of $500$ MeV and a full scale of $128$ GeV. 
As an example, electron and photon primitives are formed at L1CAL by simply applying energy thresholds to the EM energy of a single trigger tower while jet 
primitives are formed using the total EM+HAD of a single trigger tower. For electrons, tracks from the Level-1 track trigger (XFT) can be matched to the 
trigger towers while HAD energy can be used for rejection. L1CAL also calculates global SumET and MET, using the lower resolution 8-bit EM+HAD energy information. 

The main task of the existing L2CAL was to find clusters using the transverse energy ($E_T$) of trigger towers.  The cluster finding algorithm was based on a 
simple algorithm used for Run I, and was implemented in dedicated hardware. In this simple algorithm, 
the L2CAL hardware forms clusters by simply combining contiguous regions of trigger towers with non-trivial energy.  Each cluster starts with a tower above a ``seed''
threshold (typically a few GeV) and all towers above a second lower ``shoulder'' threshold that form a contiguous region with the seed tower are added to the cluster. 
The size of each cluster expands until no more shoulder towers adjacent to the cluster are found. Because of this, large ``fake clusters'' are likely to be formed as the 
occupancy of the detector increases because towers which are unrelated to any jet activity have their $E_T$ boosted above clustering thresholds. One example of such
kind of ``fake cluster'' is when towers above shoulder threshold between true jets link multiple jets together into a single large cluster (cluster merging). This would
reduce the efficiency, and increase the rate, for triggers requiring multiple jets at Level 2 at higher luminosity, such as some important triggers for Higgs and top physics.
One more limitation of the existing hardware-based L2CAL system is that it does not re-calculate SumET and MET using the full 10-bit resolution energy information 
available, instead it uses the SumET and MET information directly from L1CAL, which is based on 8-bit resolution. This design feature limits its trigger selection 
capability, or rejection power, for triggers with global transverse energy requirements.

The existing L2CAL trigger system has worked reasonably well at lower luminosity for Run II, however, 
as the occupancy in the calorimeter increases with luminosity, the simple hardware-based L2CAL system starts to lose its rejection power. 


\section{THE LEVEL 2  AND LEVEL 1 CALORIMETER TRIGGER UPGRADE}

The main goal of the upgrade is to significantly reduce the growth terms of the existing jet and MET related triggers.

The basic idea of the L2 upgrade is to provide the full $10$ bit resolution trigger tower energy information directly to the Global Level 2 decision CPU where 
a cluster finding algorithm can reconstruct jets and recalculate MET and SumET. 
The upgrade approch is based on the Pulsar board \cite{Pulsar}, a general purpose VME board developed at CDF and used for upgrading  
both the Level-2 global decision crate \cite{L2} and the Level-2 silicon vertex tracking (SVT) subsystem \cite{SVT}. 
The full resolution (10-bit) calorimeter trigger tower data are received, preprocessed and merged by a set of Pulsar boards before 
being sent to the Level 2 decision CPU where more sophisticated algorithms can be implemented. Since the actual cluster-finding is 
done inside the CPU, it is more flexible and more robust against increasing luminosity or higher occupancy in the calorimeter. 
With this approach, jet reconstruction using a cone algorithm which is currently being done at Level 3 can be moved to Level 2,
albeit clustering trigger towers (instead of physical towers) and using only a single iteration in order to save processing time.

\begin{figure}
\centering
\includegraphics[width=3.5in]{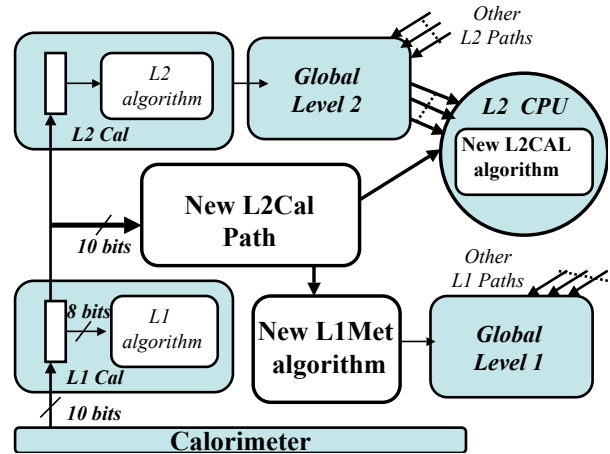}
\caption{Hardware configuration for the both L2 and L1 Trigger upgrade. The new hardware path makes available the full $10$ bits resolution trigger towers to the L2 decision CPU. The same full resolution information are sent to an additional dedicated hardware, which re-calculates the MET for the level 1.}
\label{hardware}
\end{figure}

The new system is composed by a new hardware path connecting the L1CAL directly to the L2 decision CPU (see Figure~\ref{hardware}). The same hardware has been used to transfer the trigger tower information, full 10-bits resolution, to an additional new Pulsar boards which calculates, within the L1 timing constraints, the MET for the L1 system. 

\section{PERFORMANCES}
\begin{figure}
\centering
\includegraphics[width=3.5in]{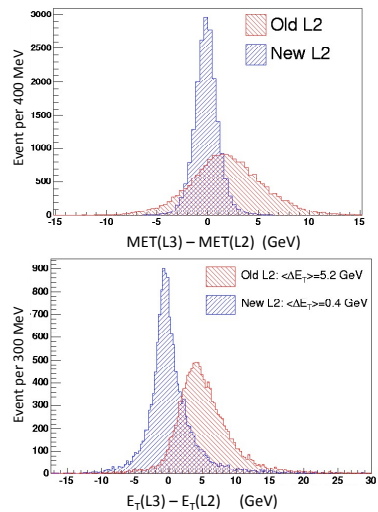}
\caption{Difference between L2 and L3 MET (top) and jet transverse energy (bottom) for existing and upgraded L2CAL system. The average luminosity is $180 \times 10^{30} cm^{-2} s^{-1}$.}
\label{efficiency}
\end{figure} 

\begin{figure}
\centering
\includegraphics[width=3.5in]{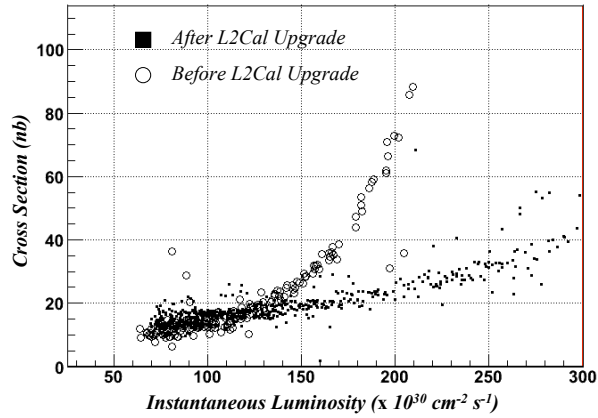}
\caption{Cross Section of the jet trigger selection requiring jets above $40$ GeV as a function of the Instantaneous Luminosity: upgraded L2CAL vs existing L2CAL}
\label{JET40}
\end{figure}


\begin{figure}
\centering
\includegraphics[width=3.5in]{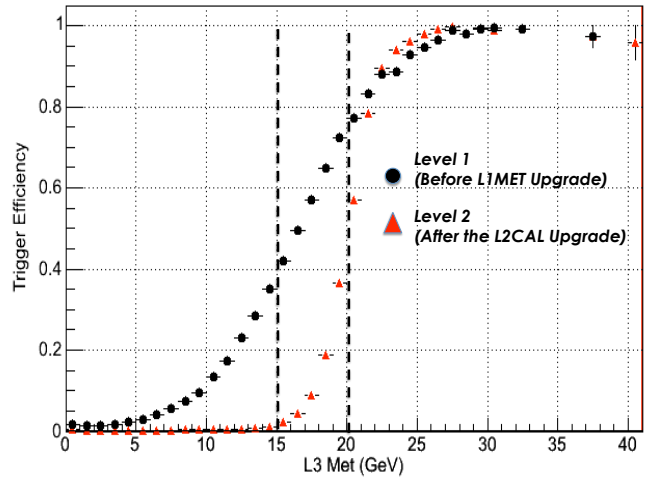}
\caption{Efficiency verse Level $3$ MET for the new Level 2 MET trigger with $20$ GeV threshold (After L2CAl Upgrade) and for the Level 1 MET trigger with $15$ GeV threshold (Before the L1MET Upgrade).}
\label{preL1MET}
\end{figure} 

\begin{figure}
\centering
\includegraphics[width=3.5in]{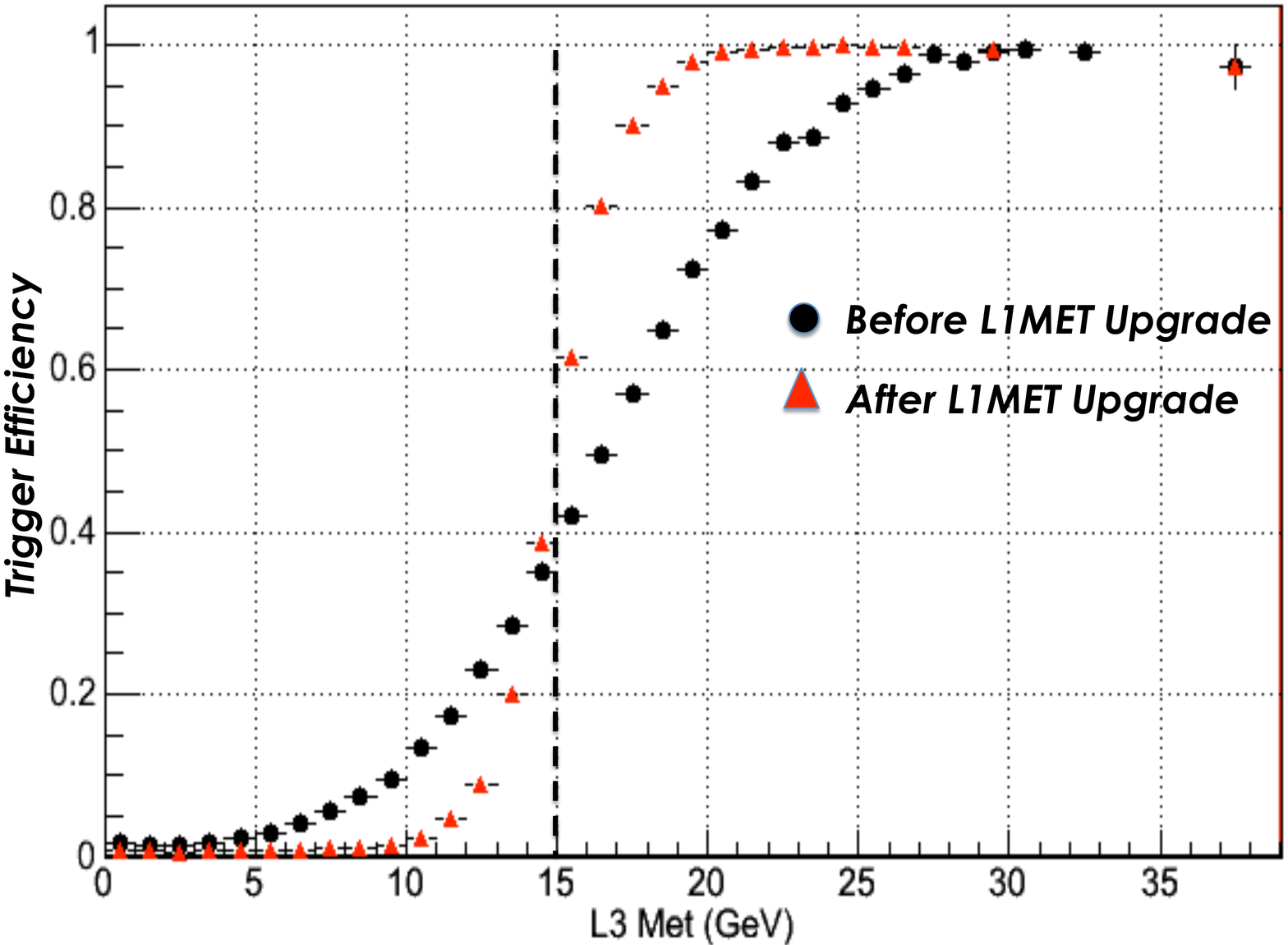}
\caption{Efficiency verse Level $3$ MET for the new Level 1 MET trigger with $15$ GeV threshold: before and after the L1MET upgrade.}
\label{afterL1MET}
\end{figure}

The Pulsar-based L2CAL upgrade has improved both jet and MET measurements at Level 2. During the design stage of the new system,  
extensive studies have already shown that Level 2 jets found using the new algorithm are nearly equivalent to offline jets in terms
of $E_T$, cluster centroid and efficiency, a vast improvement over the current situation. In addition, the calculation of MET at Level 2 with resolution
closer to that at Level 3 will significantly reduce Level 2 rates for triggers requiring MET. The upgraded L2CAL system has performed
as expected in real data taking. As an example, Figure~\ref{efficiency} shows the difference between the Level 2 and Level 3 in
MET and Jet transverse energy, for the existing system as well as for the upgraded system. 
These improvements allow a significant rate reduction as well as efficiency improvement in jet and MET based triggers.  
As examples, Figure~\ref{JET40} shows the Level 2 JET40 trigger cross section growth with luminosity before and after the upgrade. 

Several Higgs-dedicated triggers have been developed exploiting the potentialities of the L2CAL Upgrade. In particular the request of a even lower MET at Level 2 increases the acceptance on the signal. Before the L1MET upgrade the different resolution at Level 1 and Level 2 set a bottleneck to the lowest MET threshold at Level 2. Figure~\ref{preL1MET} shows the MET trigger efficiencies for the new L2CAL system with a cut of $20$ GeV  (full resolution) and the L1MET system (before the L1MET Upgrade) with a cut of $15$ GeV. Since the turn on of the Level 1 is very slow the trigger has low efficiency when instaed the Level 2 is fully efficient, causing a loosing of trigger capability. Before the L1MET Upgrade the only solution was keeping a higher threshold at Level 2.   
The L1MET Upgrade makes available at Level 1 the same MET trigger capability of the Level 2 increasing the flexibility of the MET-based triggers (see Figure~\ref{afterL1MET}).

\section{SUMMARY}

We have presented the design, the hardware and software implementation and the performance of the new calorimeter system for CDF experiment. 
The new L2CAL system makes the full resolution calorimeter trigger tower information directly available to the Level 2 decision CPU. 
The upgrade system allows more sophisticated algorithms to be implemented in software and both Level 2 jets and MET are made nearly equivalent to 
offline quality, thus significantly improving the performance and flexibility of the jet and MET related triggers. As a natural expansion of the 
upgraded Level 2 trigger system we pushed the improved MET resolution downto the Level 1, and these are big steps forward to have enough flexibility to deal with potential new challenges at the highest luminosities, and to improve CDF new physics reach sensitivities beyond baseline. 
We foresee many opportunities for additional improvements in trigger purity and efficiency, most notably for physics triggers searching for Higgs and new physics.

\end{document}